\documentclass[aps,physrev,nofootinbib,reprint]{revtex4-2}

\usepackage{amsmath,amssymb}
\usepackage{mathtools}

\usepackage[final]{graphicx}
\usepackage{relsize} 
\usepackage{bm}
\usepackage{overpic}
\usepackage[usenames,dvipsnames]{xcolor}

\usepackage{siunitx}
\ifdefined\qty\else
  \ifdefined\NewCommandCopy
    \NewCommandCopy\qty\SI
  \else
    \NewDocumentCommand\qty{O{}mm}{\SI[#1]{#2}{#3}}
  \fi
\fi
\ifdefined\unit\else
  \ifdefined\NewCommandCopy
    \NewCommandCopy\unit\si
  \else
    \NewDocumentCommand\unit{O{}m}{\si[#1]{#2}}
  \fi
\fi

\usepackage{natbib}
\usepackage{microtype}

\usepackage[colorlinks,citecolor=red,urlcolor=blue]{hyperref}
\IfFileExists{hypernat.sty}{\usepackage{hypernat}}

\newcommand{\defn}[1]{\emph{#1}}

\newcommand{\bivec}[1]{\tensor{#1}} 

\newcommand{\ccwbivec}{{\setlength{\fboxsep}{1pt}\fbox{$\circlearrowleft$}}}

\newcommand{\fourvec}[1]{\mathsf{#1}}

\newcommand{\dblcdot}{\mathbin{\vcentcolon}} 

\newcommand{\volumeVariable}{\mathcal{V}}
\newcommand{\dVol}{d\volumeVariable}

\newcommand{\surfaceVariable}{S}

\newcommand{\dvSurf}{d\vec{\surfaceVariable}}

\newcommand{\xh}{\hat{x}}
\newcommand{\yh}{\hat{y}}
\newcommand{\zh}{\hat{z}}

\begin{document}


\title{Teaching Maxwell's Equations from 2D to 3D with Bivectors}

\author{Steuard \surname{Jensen}}
\email{jensens@alma.edu} 
\author{Edward \surname{Markarian}}
\affiliation{Department of Physics and Engineering, Alma College, Alma, MI 48801}

\date{June 10, 2026}

\begin{abstract}
Electromagnetism is one of the few core physics topics without simple two-dimensional examples to start from: the cross product and curl require three dimensions. 
Previous work described magnetism as a \defn{bivector} field, visualized with oriented (clockwise/counterclockwise) ``tiles'' rather than the traditional (pseudo)vector ``arrows.''
Here, we express Maxwell's equations in this bivector language: magnetic flux is understood as a sum \emph{along} a surface rather than through it, and the magnetic field tiles \emph{encircle} the boundary of an Amp\`{e}rian loop or ribbon in a natural way.
This allows a gentle two-dimensional starting point and makes symmetry arguments natural for magnetism.
\end{abstract}

\maketitle

\section{Introduction}

Two-dimensional examples are a mainstay of physics teaching. Systems that can be analyzed in a plane are common in nature, from collisions and trusses to projectiles and orbits. They are rich enough to require vectors for analysis, but still simple to visualize and to model. They give students a chance to practice new skills before tackling more complicated three-dimensional problems.

One of the few topics in the core physics curriculum that has not usually had the benefit of simple two-dimensional examples is electromagnetism. Electrostatic fields and forces work fine in a plane, but the cross products, curls, and right-hand rules necessary for magnetism and electrodynamics inherently require three-dimensional reasoning. The Gaussian surfaces and Amp\`{e}rian loops used with Maxwell's equations are presented as three-dimensional concepts.

It turns out, though, that three-dimensional reasoning is a requirement of the traditional language for magnetism, not of magnetism itself.
If we describe magnetism as a \defn{bivector} field $\bivec{b}$ (rather than as a vector $\vec{B}$), simple two-dimensional examples are entirely natural (among other benefits).\cite{Jensen:2023mag}
As reviewed and illustrated in Section~\ref{sec:bivecReview}, a bivector quantity is 
represented by an oriented ``tile'' whose area represents its magnitude. Two-dimensional examples arise naturally: current loops create fields oriented in the plane of the loop (Fig.~\ref{fig:LoopFieldReflection}) and magnetic forces act in the bivector's plane (Fig.~\ref{fig:BivecDotProds}, left).
No right-hand rules are necessary for bivector calculations, but if desired (in three dimensions) a simple right-hand rule can relate each tile to the corresponding traditional vector.

Here, we take this bivector formalism for electromagnetism one step farther to show how an instructor can introduce Maxwell's equations first in simple two-dimensional examples before generalizing them to their full three-dimensional form.
These examples could be presented to students as warm-ups from an imagined two-dimensional universe, or more concretely as slices of three-dimensional systems with translational symmetry like an infinite line of charge or infinite solenoid. %
(The same approach generalizes to higher dimensions as well.)

\medskip

Bivectors are not a standard part of today's physics curriculum, but they can be introduced alongside (or instead of!)\ any topic in physics involving cross products. 
In previous work, the first author presented the pedagogical benefits of this approach for teaching rotational physics\cite{Jensen:2022rot} (with a student co-author) and basic magnetism\cite{Jensen:2023mag} (paralleling prior work by Jancewicz\cite{Jancewicz:1980eb}). 
Bivector magnetism avoids common student difficulties with the right-hand rule and confusion between $\vec{E}$ and $\vec{B}$, it removes the potential stumbling blocks of pseudovectors, and it works in any number of dimensions.

Given those benefits, it is natural to ask whether the bivector approach could also address common student difficulties learning Maxwell's equations. 
There is a substantial literature on specific challenges students face in this area, and even beyond the gentler two-dimensional starting point there are ways it could help. 
First, a number of authors have studied student understanding of symmetry arguments in Gauss's and Amp\`{e}re's laws (a valuable recent study by Campos et al.\cite{Campos:2023ac} discusses this, and includes a broader literature review). 
We will see that the bivector description makes symmetry arguments for magnetic fields just as natural as they are for electric fields. 
And second, students often fail to think of the integral $\int \vec{B} \cdot d\vec{\ell}$ as a sum and thus struggle with multi-leg Amp\`{e}rian loops.\cite{Wallace:2010ua}
Learning Amp\`{e}re's law first in two dimensions where this term is explicitly a sum may better prepare students to think in this way.

\medskip

This is not the only proposed reformulation of electromagnetism, but others require much more radical departures from traditional vector calculus.
The most rigorous treatments of electromagnetism use the language of differential forms. 
The recent book by Jancewicz\cite{Jancewicz:2022dq} gives a thorough, illustrated explanation of this (including a clear explanation of the important distinction between ordinary and twisted forms), and many other treatments are available (e.g.~\cite{vanDantzig:1934fe,Warnick:1997te,Fumeron:2020is,Hehl:2003fce}).
But this approach requires students to learn an entirely new algebra and calculus of forms to describe $E$ (a 1-form), $B$ (a 2-form), $D$ (a twisted 2-form Hodge-dual to $E$), and so on. Visualizing differential forms (using ``Schouten pictograms'') 
is subtle and often ill-defined,\cite{Fumeron:2020is} and discussions of forms rarely mention the pedagogically valuable concepts of Gaussian surfaces and Amp\`{e}rian loops.
For those interested in eventually learning this rigorous approach, studying the magnetic field as a bivector may provide helpful intuition for the 2-form $B$, as both are rank-2 antisymmetric tensors (contravariant and covariant, respectively).


Another proposed reformulation of electromagnetism comes in the context of geometric algebra, championed by Hestenes\cite{Hestenes:2003} and explored in detail by many others (e.g.~\cite{Doran:2003,Gull:1993,Vold:1993EM}). This formalism includes bivectors explicitly, with much the same properties as described in the present work. 
But in the geometric algebra context they are inextricably linked to a complete reformulation of the usual mathematical rules of physics. 
The present work advocates for the value of bivectors outside of geometric algebra, but the ideas here may be of interest within that subject as well.

\medskip

The organization of this paper is as follows. 
Section~\ref{sec:bivecReview} gives a brief review of the bivector description of the magnetic field. 
Section~\ref{sec:Gauss} presents Gauss's law in two and then three dimensions, and Section~\ref{sec:Ampere} does the same for Amp\`{e}re's law. Section~\ref{sec:FaradayFlux} discusses magnetic flux: Faraday's law and Gauss's law for magnetism.
Vector and bivector derivatives appear in the differential versions of Maxwell's equations in Section~\ref{sec:differentialMaxwell}.
Finally, Section~\ref{sec:conclusions} summarizes the challenges and benefits of this approach.

\section{Describing magnetism with bivectors}
\label{sec:bivecReview}

This brief overview is only intended to summarize properties of the magnetic field as a bivector that are necessary for studying Maxwell's equations. These ideas are presented in more detail in \cite{Jensen:2023mag}, though it may be helpful to first see them in the context of rotations.\cite{Jensen:2022rot}

\begin{figure}
\hspace*{\fill}
\includegraphics[width=0.4\linewidth]{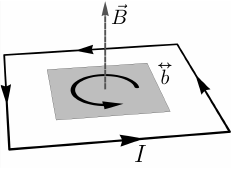}
\hfill
\hfill \hfill \hfill \hfill
\includegraphics[width=0.4\linewidth]{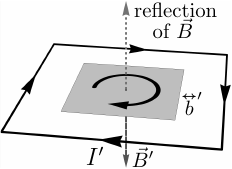}
\hspace*{\fill}
\caption{\label{fig:LoopFieldReflection} 
\textit{Left:} The magnetic field of a current loop is a \defn{bivector} quantity $\bivec{b}$, represented as a ``tile'' with an orientation (clockwise/counterclockwise). At points inside the loop, the field's orientation matches the direction of current flow. In three dimensions, the traditional magnetic field $\vec{B}$ is normal to the tile with direction related to the tile's orientation by the right-hand rule.
\textit{Right:} Under reflection, the bivector orientation naturally reverses along with the current. The corresponding vector $\vec{B}'$ must also reverse (by the right-hand rule), even though the reflection of the original $\vec{B}$ remains unchanged. This extra reversal of $\vec{B}$ under reflection is the defining characteristic of a \defn{pseudovector}.}
\end{figure}

Figure~\ref{fig:LoopFieldReflection} illustrates the essential intuition for the bivector description of magnetism: the magnetic field generated in the center of a current loop has an orientation matching the flow of current in the loop. The traditional vector magnetic field points straight through the loop: it is always normal to the corresponding bivector tile in the direction given by the right-hand rule. 

The figure also illustrates a benefit of the bivector description: under reflection, a bivector's orientation reverses exactly as the source currents do. 
The traditional vector magnetic field is not so natural: after reflecting the coordinates, you must also reverse the direction of $\vec{B}$ (essentially because reflection changes a right hand into a left hand).
This is the defining characteristic of a \defn{pseudovector} (or \defn{axial vector}) quantity rather than an ordinary (\defn{polar}) vector quantity, but it is rare for teachers and textbooks to directly address that distinction.

\begin{figure}
\hspace*{\fill}
\raisebox{-0.5\height}{\includegraphics[width=4cm]{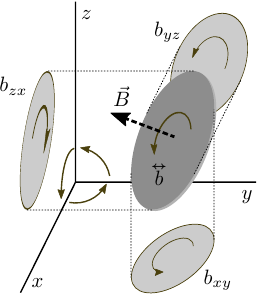}}
\hfill 
\hfill \hfill \hfill \hfill \hfill \hfill \hfill
\raisebox{-0.5\height}{\includegraphics[width=3.5cm]{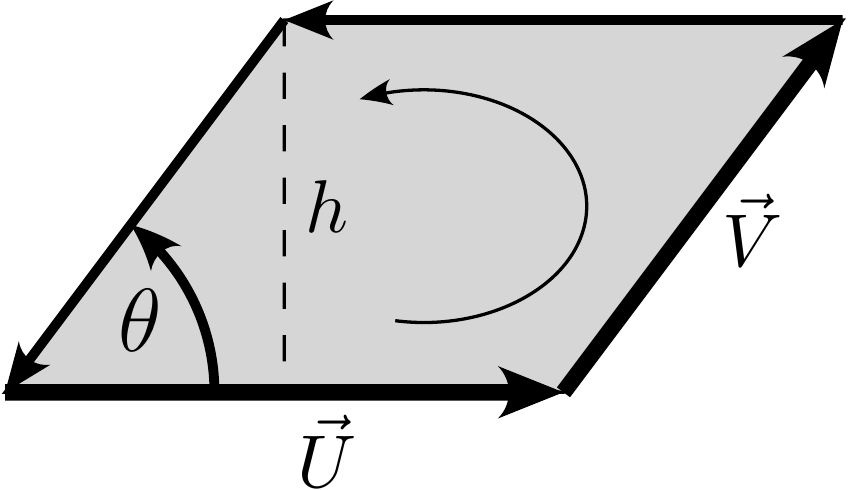}}
\hspace*{\fill}
\caption{\label{fig:TileComponentsWedge} 
\textit{Left:} The components of a bivector $\bivec{b}$ can be found by projecting its tile onto each of the three coordinate planes.
In this example, the projected orientations show that $b_{xy}$ and $b_{yz}$ are positive while $b_{zx}$ is negative.
\textit{Right:} The wedge product $\bivec{b} = \vec{U} \wedge \vec{V}$ of two vectors is visualized as a parallelogram tile of area $|\bivec{b}| = |\vec{U}| h = |\vec{U}| |\vec{V}| \sin\theta$. Its orientation follows the first vector tip-to-tail with the second (in order).}
\end{figure}

Just as a vector's components describe the arrow's projected length onto each of the coordinate axes, the coordinate components of a bivector describe the tile's projected area onto each of the coordinate planes, as shown in Fig.~\ref{fig:TileComponentsWedge} (left): $b_{zx}$ is how much the tile aligns with the $zx$-plane, with a positive sign matching the orientation that rotates $+z$ toward $+x$. 
Since $b_{xz}$ means the same thing but with positive meaning the orientation that rotates $+x$ toward $+z$, we can see that $b_{xz} = - b_{zx}$.
Overall, the components form an antisymmetric matrix: 
\begin{align}
\label{eq:bivecComponentIndices}
\bivec{b} &= 
\begin{pmatrix}
 0 & b_{xy} & -b_{zx} \\
-b_{xy} & 0  & b_{yz} \\
b_{zx} & -b_{yz} & 0
\end{pmatrix}
=
\begin{pmatrix}
 0 & B_z & -B_y \\
-B_z & 0  & B_x \\
B_y & -B_x & 0
\end{pmatrix}
\;.
\end{align}
The second form shows the correspondence between the bivector and traditional pseudovector components.

The \defn{wedge product} of two vectors produces a bivector: $\bivec{b} = \vec{U} \wedge \vec{V}$. As illustrated in Fig.~\ref{fig:TileComponentsWedge} (right), geometrically we place $\vec{U}$ and $\vec{V}$ tip-to-tail (in order!) and the resulting parallelogram is the bivector tile. 
(Reversing the order reverses the bivector's orientation.)
The formula for parallelogram area, $|\bivec{b}| = |\vec{U}| h = |\vec{U}| |\vec{V}| \sin\theta$, matches the magnitude of the traditional cross product: $\vec{U} \times \vec{V}$ is related to this bivector by the right-hand rule. In components,
\begin{equation}
b_{ij} = U_i V_j - U_j V_i
\quad,
\end{equation}
matching the usual cross product components.%
\footnote{We use index notation here only for spacelike Cartesian coordinates, so for simplicity all tensor indices are shown as subscripts.}

Just as a vector can be written in terms of unit vectors, $\vec{U} = U_x \,\xh + U_y \,\yh + U_z \,\zh$, a bivector can be written as
\begin{equation}
\bivec{b} = b_{xy} \,\xh\wedge\yh + b_{yz} \,\yh\wedge\zh + b_{zx} \,\zh\wedge\xh
\quad.
\end{equation}
(There is no need for (e.g.)\ a separate $b_{yx}$ term, because $\xh\wedge\yh$ is already antisymmetric.)
When the $xy$-plane is drawn in the plane of the paper, we sometimes use $\ccwbivec$ as shorthand for $\hat{x}\wedge\hat{y}$.

\begin{figure}
\hspace*{\fill}
\raisebox{-0.5\height}{\includegraphics[width=3cm]{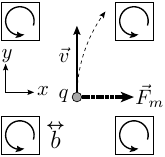}}
\hfill
\hfill \hfill \hfill \hfill \hfill \hfill \hfill
\raisebox{-0.5\height}{\includegraphics[width=4cm]{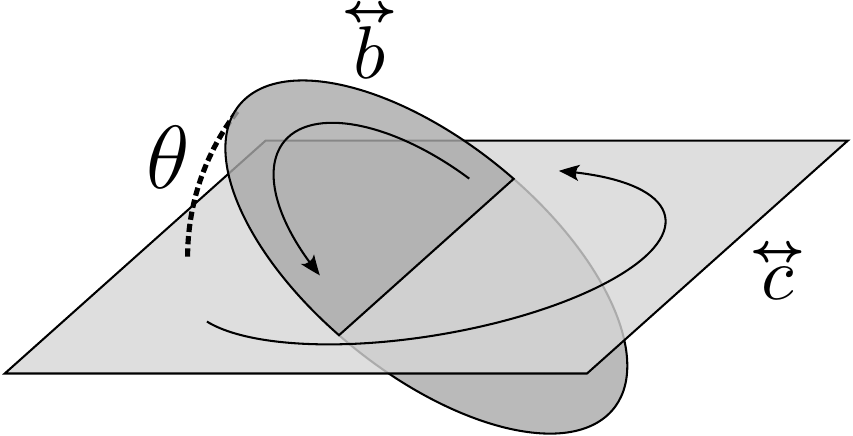}}
\hspace*{\fill}
\caption{\label{fig:BivecDotProds} 
\textit{Left:} To find the direction of the magnetic force $\vec{F} = q\, \bivec{b} \cdot \vec{v}$ on a particle of charge $+q$ moving with velocity $\vec{v}$ in a uniform magnetic field, first project the velocity into the plane of $\bivec{b}$ and then rotate it by \qty{90}{\degree} \emph{opposite} the field's orientation.
\textit{Right:} The \defn{double dot product} is the scalar product of two bivectors $\bivec{b}$ and $\bivec{c}$: the product of their same-directed magnitudes. Its value is $\frac{1}{2} \bivec{b}\dblcdot\bivec{c} = \frac{1}{2} b_{ij} c_{ij} = |\bivec{b}|\,|\bivec{c}|\,\cos\theta$.}
\end{figure}

The magnetic force on a point charge $q$ with velocity $\vec{v}$ is $\vec{F}_m = q\, \bivec{b} \cdot \vec{v}$. The dot product denotes tensor index contraction of adjacent indices $F_{m,i} = q\, b_{ij} v_j$ (using Einstein summation notation of repeated indices):
this is equivalent to matrix multiplication with column matrix $\vec{v}$. 
Because $b_{ij}$ is antisymmetric, this dot product is also antisymmetric:
$F_{m,i} = -q\, v_j b_{ji}$, so $\vec{F}_m = -q\, \vec{v} \cdot \bivec{b}$.

Geometrically, this is illustrated in Fig.~\ref{fig:BivecDotProds} (left): the force's direction is found by projecting the velocity vector into the plane of the bivector and then rotating it \qty{90}{\degree} \emph{opposite} the bivector orientation. The resulting magnitude is $|\vec{F}_m| = |q| \, |\bivec{b}| \, |\vec{v}| \cos\theta$, where $\theta$ is the angle between the velocity and the bivector plane.%
\footnote{This $\cos\theta$ does agree with the $\sin\theta$ familiar from $q\, \vec{v} \times \vec{B}$, since the angle between the velocity and the normal vector $\vec{B}$ is $(\qty{90}{\degree} - \theta)$.}

Finally, Maxwell's equations will sometimes require the scalar product of two bivectors, denoted by a ``double dot product'' corresponding to tensor index contraction of both indices: $\bivec{b} \dblcdot \bivec{c} = b_{ij} c_{ij}$, as illustrated in Fig.~\ref{fig:BivecDotProds} (right).
This plays the same role as the dot product $\vec{B} \cdot \vec{C}$ of the corresponding pseudovectors, and the results are the same.
For example, the magnitude of a bivector obeys $|\bivec{b}|^2 = \tfrac{1}{2} \bivec{b} \dblcdot \bivec{b} = b_{xy}^2 + b_{yz}^2 + b_{zx}^2$.
(The factor of $\tfrac{1}{2}$ compensates for the double counting of the antisymmetric bivector components.)
More generally, 
$\frac{1}{2} \bivec{b} \dblcdot \bivec{c} 
= |\bivec{b}|\,|\bivec{c}|\,\cos\theta$,
where $\theta$ is the angle between the tiles.

\section{Gauss's law in 2D and 3D}
\label{sec:Gauss}

There are no bivectors involved in Gauss's law for electric fields.
For a two-dimensional region $\mathcal{A}$ whose boundary is the curve $\mathcal{C}$ (the ``Gaussian surface'' curve), the flux of the electric field $\vec{E}$ through the surface must equal the total charge enclosed:
\begin{equation}
\label{eq:Gauss2D}
\oint_{\mathcal{C}} \vec{E} \cdot \dvSurf = Q_\text{encl}/\epsilon_0
 = \frac{1}{\epsilon_0} \int_\mathcal{A} \sigma \,da
 \quad.
\end{equation}
Here, the surface vector is $\dvSurf = d\ell\, \hat{n}$, where $d\ell$ is an infinitesimal element of the surface's length and $\hat{n}$ is its outward normal vector.%
\footnote{Throughout this discussion, the term ``surface'' (as in ``surface area'' or ``surface curve'') and 
symbol $\dvSurf$ refer to a surface that a vector field or current can flow across: a $(d-1)$-dimensional surface in $d$-dimensional space.}
The second equality expresses the enclosed charge as a sum of area charge density %
$\sigma$. 

\begin{figure}
\centering
\includegraphics{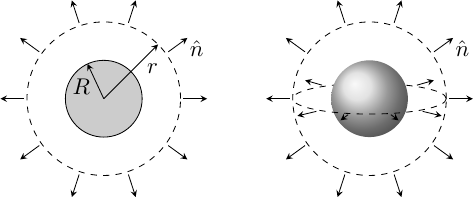} 
\caption{\label{fig:GaussCircleSphere}%
\textit{Left:}
In two dimensions, an imagined Gaussian ``surface'' of radius $r$ (dashed line) can be used to find the electric field of a circular region of radius $R$ with constant charge density. Normal vectors $\hat{n}$ to the surface are shown.
\textit{Right:}
A similar spherical charge distribution in three dimensions.%
}
\end{figure}

As usual, Gauss's law is most useful for computing electric fields for symmetric charge distributions. Figure~\ref{fig:GaussCircleSphere} (left) illustrates this for a circular region of radius $R$ with uniform charge density $\sigma$. (We can also interpret this as a three-dimensional example: a horizontal slice of an infinite straight uniformly charged cylinder.) The process begins with a symmetry argument just as in three dimensions: in polar coordinates $(r,\theta)$, the charges' rotational symmetry guarantees that $\vec{E}$ is independent of $\theta$, and reflectional symmetry across any line through the origin forbids a $\hat{\theta}$ component. Thus, $\vec{E}(r,\theta) = E(r) \, \hat{r}$.

To apply Gauss's law, we choose a circular Gaussian surface $\mathcal{C}$ of radius $r$ whose normal vector is always $\hat{n} = \hat{r}$, so $\vec{E} \cdot \hat{n} = E(r)$. The left side of Gauss's law is then straightforward to evaluate:
\begin{equation}
\oint_{\mathcal{C}} \vec{E} \cdot \dvSurf
  = \oint_{\mathcal{C}} E(r) \,d\ell = E(r) \oint_{\mathcal{C}} d\ell 
  = 2\pi r \, E(r)
 \quad.
\end{equation}
Meanwhile, the enclosed charge is charge density times an appropriate area:
either $Q_\text{encl} = Q \equiv \pi R^2 \sigma$ if $r>R$ so the whole charged area is enclosed, or $Q_\text{encl} = \pi r^2 \sigma = \frac{r^2}{R^2} Q$ if $r<R$.

Solving for the electric field, this leads to
\begin{equation}
\vec{E}(r) = \begin{cases}
\frac{Q}{2\pi \epsilon_0 \, r}  \,\hat{r}, & r \ge R
\\
\frac{Q \, r}{2\pi \epsilon_0 \, R^2} \, \hat{r}, & r \le R
\end{cases}
\quad.
\end{equation}
This matches the familiar result for the electric field of an infinite cylinder in three dimensions: the only difference from the usual procedure was that our surface was a circle rather than a cylinder (of arbitrary length).

\subsection{Extending Gauss's law to three dimensions}

The transition to the three-dimensional version of Gauss's law is straightforward and entirely matches standard treatments. The flux of the electric field now passes through a two-dimensional (Gaussian) surface area $\mathcal{A}$ that is the boundary of a three-dimensional volume $\mathcal{V}$:
\begin{equation}
\label{eq:Gauss3D}
\oint_{\mathcal{A}} \vec{E} \cdot \dvSurf = Q_\text{encl}/\epsilon_0
 = \frac{1}{\epsilon_0} \int_\mathcal{V} \rho \, \dVol
 \quad.
\end{equation}
Here, the surface vector is $\dvSurf = da \, \hat{n}$, where $da$ is an infinitesimal element of the surface's area and $\hat{n}$ is again its outward normal vector.

Figure~\ref{fig:GaussCircleSphere} (right) shows a spherical charge distribution surrounded by a spherical Gaussian surface (along with some normal vectors): the analysis in this case is exactly as typically taught.
As mentioned earlier, the left image in that figure (with a circular charge) could also be interpreted as an end-on view of an infinite charged cylinder: in that case, the circular Gaussian surface in two dimensions is similarly extended to a cylindrical surface (just as in traditional treatments of such systems).

\section{Amp\`{e}re's law in 2D and 3D}
\label{sec:Ampere}

The next Maxwell equation usually encountered is Amp\`{e}re's law for magnetostatic current distributions.
This takes a somewhat different form for bivector magnetism, especially in two dimensions.
Rather than an Amp\`{e}rian loop as in three dimensions, in two dimensions we choose an ``Amp\`{e}rian ribbon:'' an arbitrary imaginary one-dimensional surface curve $\mathcal{C}$ that current can flow through (like runners crossing a finish line) with a chosen positive direction of flow. 
The boundary 
of the surface curve $\mathcal{C}$  consists of 
its two endpoints $p$. 

Amp\`{e}re's law says that
the total \defn{encirclement} of the boundary (explained below) by the magnetic bivector field $\bivec{b}$ equals the total current flowing through the surface:
\begin{equation}
\label{eq:Ampere2D}
\sum_{p} 
\tfrac{1}{2} \bivec{b} \dblcdot \bivec{c}_p
  = \mu_0 I
  = \mu_0 \int_\mathcal{C} \vec{K} \cdot \dvSurf
\quad.
\end{equation}
The new element here is the \defn{encircling (unit) bivector} $\bivec{c}_p$ around (at) each point $p$: a unit bivector that defines the orientation on the boundary in much the same way that a normal vector $\hat{n}$ defines an orientation of a surface. 
The final integral writes the current $I$ passing through the surface curve $\mathcal{C}$ as an integral of area current density %
$\vec{K}$ (with SI units \unit{\A\per\m}).
As in the two-dimensional Gauss's law, the surface vector is $\dvSurf = d\ell\, \hat{n}$.

\bigskip

\begin{figure}
\centering
\includegraphics{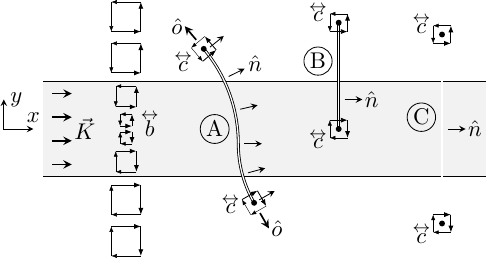} 
\caption{\label{fig:AmpereLine2D}%
A uniform current density $\vec{K}$ flows along a straight wire in two dimensions, with the magnetic field bivector $\bivec{b}$ clockwise below the wire and counterclockwise above, vanishing to zero in the middle.
In two dimensions, Amp\`{e}re's law relates current flow through an imagined ``Amp\`{e}rian ribbon'' (like crossing a finish line) to the \defn{encirclement} of the ribbon's boundary points by the magnetic field.
Each Amp\`{e}rian ribbon shown (A, B, and C) is labeled with typical normal vectors (in the chosen positive direction of flow), and ribbon~A also labels the ``outward vector'' $\hat{o}$ at each endpoint. Each of the ribbons' endpoints is labeled with its own \defn{encircling bivector} $\bivec{c} = \hat{n} \wedge \hat{o}$. This definition ensures that positive encirclement crosses the ribbon in the same direction chosen as positive $\hat{n}$, like a ``water wheel'' pushed by positive current flow inside.}
\end{figure}

In parallel to our discussion of Gauss's law, we will make this concrete by using Amp\`{e}re's law to compute the magnetic field for current distributions with substantial symmetry.
Figure~\ref{fig:AmpereLine2D} illustrates (part of) an infinite straight wire in two dimensions ($x$ and $y$, as shown), carrying a uniform area current density $\vec{K}$. (Interpreted as a slice of a three-dimensional example, this would be uniform current through an infinite conducting slab.)

We again begin by using symmetry to constrain the form of the magnetic field.
The only possible orientations for a bivector in 2D are clockwise or counterclockwise in the plane.
Translational symmetry of the current distribution guarantees that $\bivec{b}$ is independent of $x$, and symmetry under vertical reflections (across the $x$-axis) shows that it must have opposite orientation above and below the center of the wire, and thus that the field must be zero along the axis. We can write $\bivec{b} = b(y) \, \hat{x}\wedge\hat{y} = b(y) \, \ccwbivec$ (or in component form, $b_{xy} = b(y)$).

It's worth taking a moment to observe that symmetry arguments like this are just as easy for bivector magnetic fields as they are for vector electric fields. 
This is not true for the traditional pseudovector description of magnetism, where teachers and textbooks tend to silently ignore reflectional symmetry arguments (and hope that students don't ask about them) to avoid discussing the potentially confusing reversal of $\vec{B}$ under reflection.


We next choose our Amp\`{e}rian ribbon and specify its positive direction of flow (corresponding to a normal vector $\hat{n}$ at each point). Each endpoint of the ribbon also has an ``outward unit vector'' $\hat{o}$ pointing away from the endpoint in the ribbon's direction (as illustrated for ribbon~A in Fig.~\ref{fig:AmpereLine2D}). 
These two unit vectors together define a bivector orientation encircling each endpoint: the encircling (unit) bivector $\bivec{c} = \hat{n} \wedge \hat{o}$. 
The orientations of the two ends are opposite one another regardless of the shape of the ribbon. Because the bivector tile $\bivec{c}$ is constructed by placing $\hat{n}$ and $\hat{o}$ tip-to-tail in that order, the ``inner'' side of the tile (that crosses the ribbon) always circulates through in the same (positive) direction as $\hat{n}$.

For the scenario in Fig.~\ref{fig:AmpereLine2D}, the simplest choice of Amp\`{e}rian ribbon is a vertical line from the center ($y=0$) to an arbitrary height $y$ with the positive direction chosen as $+x$, shown in the figure as ribbon~B. 
The magnetic field at the $y=0$ endpoint is zero (by symmetry), so only the top endpoint contributes to the left side of Eq.~\eqref{eq:Ampere2D}. At the top endpoint, $\hat{n}=\hat{x}$ and $\hat{o} = \hat{y}$, so the encircling bivector is $\bivec{c} = \hat{x}\wedge\hat{y} = \ccwbivec$.
The left side of Eq.~\eqref{eq:Ampere2D} is just %
$\frac{1}{2} \bivec{b} \dblcdot \bivec{c} = \frac{1}{2} b(y)\, (\hat{x}\wedge\hat{y})\dblcdot(\hat{x}\wedge\hat{y}) = b(y)$.

Meanwhile, the current flows parallel to the ribbon's normal vector, so the right side of Eq.~\eqref{eq:Ampere2D} becomes simple multiplication. If the wire's total width is $w$, then for any $y \ge w/2$ half of the total current $I_\text{tot}$ crosses the ribbon: $I = K \, w/2= I_\text{tot}/2$. But if $0\le y \le w/2$ then $I = K \, y = \frac{y}{w} I_\text{tot}$.
Assembling all these results into Amp\`{e}re's law (and including symmetry) gives the result
\begin{equation}
\bivec{b}(y) = \begin{cases}
\phantom{-{}}\mu_0 I_\text{tot}/2 \;\; \hat{x}\wedge\hat{y}, & y \ge w/2
\\
\mu_0 I_\text{tot} y/w \;\;  \hat{x}\wedge\hat{y}, & -w/2 \le y \le w/2
\\
-\mu_0 I_\text{tot}/2 \;\; \hat{x}\wedge\hat{y}, & y \le -w/2
\end{cases}
\quad.
\end{equation}
The negative values when $y<0$ correspond to opposite field orientation, as expected from symmetry.

Ribbon~C in Fig.~\ref{fig:AmpereLine2D} (which extends symmetrically between $\pm y$) could be used to find the same result with more symmetry but a little more work. The right side of Amp\`{e}re's law now includes both sides of the current, doubling that result. The left side now has both endpoints at nonzero $\bivec{b}$, but their oppositely oriented encircling bivectors $\bivec{c}$ exactly match the opposite directions of the magnetic field expected by symmetry, so the left side is doubled as well and the solution remains the same.

\medskip

\begin{figure}
\centering
\includegraphics{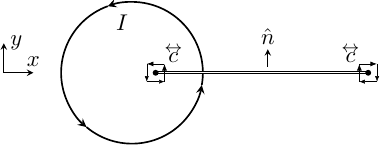} 
\caption{\label{fig:AmpereLoop2D}%
A current $I$ flows around a loop in two-dimensional space. 
An imagined Amp\`{e}rian ribbon with chosen normal vector $\hat{n}$ and corresponding endpoint encircling bivectors $\bivec{c}$ can be used to compute the magnetic field.} 
\end{figure}

Amp\`{e}re's law can also be used to compute the magnetic field due to a current loop in two dimensions (corresponding to an infinite solenoid in three dimensions). Figure~\ref{fig:AmpereLoop2D} shows a circular loop, but the same reasoning applies for any closed shape.
Choose an Amp\`{e}rian ribbon with one endpoint at any point inside the loop and the other at any point outside the loop: the total current crossing the ribbon is always $I$ regardless of the point chosen.

For a finite current distribution the field must go to zero at infinity even in two dimensions. 
If the ribbon extends to infinity then $\bivec{b} \dblcdot \bivec{c}=0$ at the outside endpoint, and since moving that endpoint leaves the rest of Amp\`{e}re's law unchanged, $\bivec{b}$ must be uniform and equal to zero at all outside points. Then for any point inside the loop, Eq.~\eqref{eq:Ampere2D} becomes
$\frac{1}{2} \bivec{b} \dblcdot \bivec{c} = \frac{1}{2} \bivec{b} \dblcdot (\hat{x}\wedge\hat{y}) = \mu_0 I$.
So the magnetic field inside is uniform, and if $I$ crosses the ribbon in the positive direction then the field's bivector orientation matches the encircling orientation of the endpoint: $\bivec{b} = \mu_0 I \, (\hat{x}\wedge\hat{y}) = \mu_0 I \,\ccwbivec$.

\medskip

Adding Maxwell's modification term involving the displacement current does not change the methods and reasoning described here in any novel way:
\begin{equation}
\label{eq:AmpereMaxwell2D}
\sum_{p} 
\tfrac{1}{2} \bivec{b} \dblcdot \bivec{c}_p
  = \mu_0 (I + I_{d})
  = \mu_0 \int_\mathcal{C} \left(\vec{K} + \epsilon_0 \frac{\partial \vec{E}}{\partial t} \right) \cdot \dvSurf
\;.
\end{equation}

\subsection{Extending Amp\`{e}re's law to three dimensions}

The bivector version of Amp\`{e}re's law in three dimensions may look a little unfamiliar at first, but it emerges naturally as we transition from two to three dimensions.
We begin from the scenario in Fig.~\ref{fig:AmpereLine2D}, where current flows like water along a two-dimensional ``channel'' and is measured as it crosses Amp\`{e}rian ``ribbon'' C. The boundary endpoints of that ribbon have bivector orientations, like water wheels that ``spin'' along with the crossing current.

\begin{figure}
\centering
\includegraphics{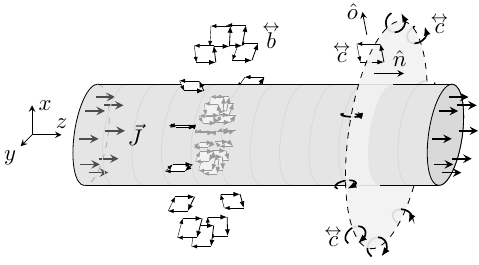} 
\caption{\label{fig:AmpereLine3D}%
A uniform current flows through a straight cylindrical wire in three dimensions.
The magnetic field bivector $\bivec{b}$ at any point lies in a plane containing the axis of the wire; its orientation on the ``inside edge'' (nearest the axis) matches the current's direction.
In going from two to three dimensions, the previous image of a flat ``channel'' of current crossing a ``ribbon'' becomes a round ``pipe'' of current crossing a ``net:'' a two-dimensional surface bounded by a traditional Amp\`{e}rian loop.
Amp\`{e}re's law relates current flow through this surface (shown with a typical normal vector $\hat{n}$) to the total \defn{encirclement} by the magnetic field summed around the boundary loop.
 Each point on the boundary can be labeled with its own \defn{encircling bivector} $\bivec{c} = \hat{n} \wedge \hat{o}$, where $\hat{o}$ is the ``outward vector'' at that point (shown explicitly for one point). This definition ensures that the positive orientation circulates through the surface in the same direction chosen for $\hat{n}$, again like a ``water wheel.''}
\end{figure}

Now expand into three dimensions. Imagine that flowing current as just the central plane of a ``pipe'' as shown in Fig.~\ref{fig:AmpereLine3D}.
Instead of crossing a one-dimensional ribbon, this flowing current is measured as it crosses a two-dimensional surface ``net.''
The ribbon's zero-dimensional boundary points extend to a familiar one-dimensional Amp\`{e}rian loop around the boundary of the net. 
Each point on the loop has its own bivector orientation that can still be pictured as a water wheel that spins along with the inner current and is still given by the encircling (unit) bivector $\bivec{c}$ at that point. As before, this is defined in terms of the normal vector and the outward unit vector: $\bivec{c} = \hat{n} \wedge \hat{o}$.

In this formulation, Amp\`{e}re's law applied to an imagined surface $\mathcal{A}$ with boundary curve $\mathcal{C}$ is written
\begin{equation}
\label{eq:AmpereMaxwell3D}
\oint_\mathcal{C} \tfrac{1}{2} \bivec{b} \dblcdot d\bivec{P}
  = \mu_0 (I + I_{d})
  = \mu_0 \int_\mathcal{A} 
    \left( \vec{J} + \epsilon_0 \frac{\partial \vec{E}}{\partial t} \right)  \cdot \dvSurf
\;.
\end{equation}
The curve is split into infinitesimal path elements $d\bivec{P} = d\ell \, \bivec{c}$ (each with its encircling bivector orientation normal to the line of the curve, like beads on a string).
The integral on the left adds up how much the bivector magnetic field matches the orientation of the path at each point (exactly like the finite sum in two dimensions).
This equals the traditional $\oint_\mathcal{C} \vec{B} \cdot d\vec{\ell}$, but with more vivid intuition: encircled current ``spins the magnetic water wheels.''

\bigskip

Using Amp\`{e}re's law to compute the magnetic field for an infinite cylindrical wire (as in Fig.~\ref{fig:AmpereLine3D}) gives another opportunity to highlight how easily symmetry arguments can simplify calculations in the bivector formalism. 
When a current distribution has reflectional symmetry across a particular plane, then at any point in that plane the only bivectors left unchanged by reflection are those whose orientations are parallel to the plane itself.%
\footnote{An equivalent ``mirror rule'' for pseudovector $\vec{B}$ is given in \cite{Moore:SI4E}, justified via the Biot-Savart law rather than symmetry.}
(This type of argument would be confusing and frustrating for pseudovector $\vec{B}$: the directions left unchanged under reflection are exactly the ones that are \emph{forbidden} for $\vec{B}$.)

In Fig.~\ref{fig:AmpereLine3D}, the current distribution has reflectional symmetry across \emph{any} plane containing the central axis. Since every point in space lies on such a plane, the field $\bivec{b}$ throughout space must be oriented along those planes, tilted toward the axis: proportional to $\hat{z} \wedge \hat{r}$ in cylindrical coordinates $(r,\theta,z)$. Translational symmetry implies $\bivec{b}$ is independent of $z$ and rotational symmetry implies it is independent of $\theta$. Altogether, $\bivec{b} = b(r) \, \hat{z} \wedge \hat{r}$. 

The calculation of Eq.~\eqref{eq:AmpereMaxwell3D} for a wire of radius $R$ proceeds from this point almost exactly as in standard treatments. The normal vector throughout the Amp\`{e}rian loop's surface is $\hat{z}$ and the outward unit vector at any boundary point is $\hat{r}$, so $\bivec{c} = \hat{z} \wedge \hat{r}$ and the left integral is
\begin{multline}
\oint_\mathcal{C} \tfrac{1}{2} \bivec{b} \dblcdot d\bivec{P}
 = \oint_\mathcal{C} \tfrac{1}{2} (b(r) \, \hat{z} \wedge \hat{r})
   \dblcdot (d\ell \, \hat{z} \wedge \hat{r})
\\
 = b(r) \oint_\mathcal{C} d\ell
 = 2\pi r \, b(r)
\;.
\end{multline}
The encircled current is $I_\text{tot}$ if $r>R$ and 
$\frac{\pi r^2}{\pi R^2} I_\text{tot}$ 
if $r<R$, so the final result for the magnetic field is
\begin{equation}
\bivec{b}(r) = \begin{cases}
\phantom{-{}}\mu_0 I_\text{tot}/2\pi r \;\; \hat{z}\wedge\hat{r}, & r \ge R
\\
\mu_0 I_\text{tot} r/2\pi R^2 \;  \hat{z}\wedge\hat{r}, & 0 \le r \le R
\end{cases}
\quad.
\end{equation}

\bigskip

Similar reasoning applies to other familiar current distributions in three dimensions. For example, if Fig.~\ref{fig:AmpereLoop2D} is interpreted as an end-on view of an infinite solenoid, the current distribution has reflectional symmetry across the plane of the page: this guarantees that the bivector field $\bivec{b}$ is parallel to the plane.
Extending the Amp\`{e}rian ribbon shown there to the traditional rectangular loop (into the page) leads to a calculation similar to those in standard treatments. (The encirclement integral of $\bivec{b}$ along the radial loop segments is zero because their encircling unit vectors $\bivec{c}$ are perpendicular to the plane of $\bivec{b}$.)

\section{Faraday's law and Gauss's law for magnetism}
\label{sec:FaradayFlux}

The final two Maxwell equations involve integrals of the magnetic field over a two-dimensional area $\mathcal{A}$. 
Following tradition, we
will still call this the ``magnetic flux'' $\Phi$:
\begin{equation}
\label{eq:BivecFlux}
\Phi \equiv \int_\mathcal{A} \tfrac{1}{2} \bivec{b} \dblcdot d\bivec{a}
\quad.
\end{equation}
This is mathematically equal to the traditional pseudovector flux $\int_\mathcal{A} \vec{B} \cdot d\vec{a}$,
but the bivector field is integrated \emph{along} the area $\mathcal{A}$: nothing is passing \emph{through} the area from this perspective.
Unlike the previous sections, this definition and the Maxwell equations it appears in are the same regardless of the number of dimensions of space.

\begin{figure}
\centering
\includegraphics{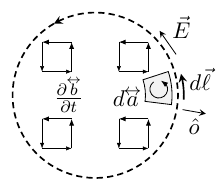} 
\caption{\label{fig:FaradayLoop2D}%
To illustrate Faraday's law, an imagined loop encircles a changing magnetic field $\bivec{b}$ that is integrated over encircled area elements $d\bivec{a}$. An electric field $\vec{E}$ is integrated along length elements $d\vec{\ell}$ of the boundary. The (bivector) orientation of the area must match the (vector) orientation of the boundary. 
(The outward vector $\hat{o}$ relates the two: $d\vec{\ell}$ points in the $\hat{o} \cdot d\bivec{a}$ direction, or equivalently, the orientation of $d\bivec{a}$ matches $\hat{o} \wedge d\vec{\ell}$.)} 
\end{figure}

From the bivector perspective, the two sides of Faraday's law have the same basic character: a field integrated \emph{along} a region of matching dimension.
\begin{equation}
\label{eq:Faraday}
\oint_\mathcal{C} \vec{E} \cdot d\vec{\ell} 
  = - \frac{\partial \Phi}{\partial t}
  \equiv - \frac{\partial}{\partial t} \int_\mathcal{A} \tfrac{1}{2} \bivec{b} \dblcdot d\bivec{a}
\quad.
\end{equation}
On the right, the magnetic field bivector (represented by a two-dimensional tile) is integrated over bivector area elements $d\bivec{a}$ oriented along a two-dimensional (but not necessarily flat) area $\mathcal{A}$. 
On the left, the electric field vector (represented by a one-dimensional arrow) is integrated over vector length elements $d\vec{\ell}$ pointing along the one-dimensional boundary curve $\mathcal{C}$ of that area. 

A typical scenario involving Faraday's law is illustrated in Fig.~\ref{fig:FaradayLoop2D}.
The bivector orientation of the area $\mathcal{A}$ must be consistent with the orientation of the vector $d\vec{\ell}$ around the boundary $\mathcal{C}$: visually, they must circle in the same direction.
(Formally, the outward unit vector $\hat{o}$ relates the two: $d\vec{\ell}$ points in the $\hat{o} \cdot d\bivec{a}$ direction, or equivalently, the orientation of $d\bivec{a}$ matches $\hat{o} \wedge d\vec{\ell}$.)
On a practical level, there is nothing ``new'' here over traditional presentations of Faraday's law. The only difference is that the bivector definition of magnetic flux is well-defined in any number of dimensions.

\bigskip

The fourth and final Maxwell equation is Gauss's law for magnetism, 
\begin{equation}
\label{eq:GaussMag}
\Phi_\text{closed} = \oint_\mathcal{A} \tfrac{1}{2} \bivec{b} \dblcdot d\bivec{a}
  = 0
\quad.
\end{equation}
That is, the magnetic flux along a \emph{closed} two-dimensional area (typically, the boundary of a three-dimensional volume) must sum to zero.
As seen from inside the volume, the amounts of ``clockwise'' and ``counterclockwise'' magnetic field on the boundary must balance out.

\begin{figure}
\centering
\includegraphics{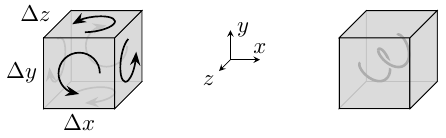} 
\caption{\label{fig:MagneticGauss}%
\textit{Left:} A closed surface enclosing a three-dimensional volume. Each face has a bivector orientation, chosen by convention to look counterclockwise when seen from outside.
\textit{Right:} An equivalent convention is to consider a right-handed helix inside the volume: follow the helix axis toward the surface and the rotation should match the surface's bivector orientation. 
(Magnetic monopoles would carry ``handedness'' charge.)}
\end{figure}

Notably, for Maxwell's equations in two-dimensional space this simply does not apply: a finite area must always have a boundary edge, so this equation is not relevant. 
In three dimensions, integration over a surface like the one in Figure~\ref{fig:MagneticGauss} is equivalent for practical purposes to the traditional treatment of magnetic flux, just with a slightly different dot product.%
\footnote{If magnetic monopoles were to exist, their charge would be a \defn{trivector} quantity (or ``pseudoscalar'', in 3D): they are not positive or negative, but rather right-handed or left-handed.}

\section{Differential versions of Maxwell's equations}
\label{sec:differentialMaxwell}

The notation in the differential versions of Maxwell's equations will require some explanation:
\begin{align}
\label{eq:DifferentialGauss}
&\text{Gauss's law:}&
\nabla \cdot \vec{E} &= \rho / \epsilon_0
 \\
\label{eq:DifferentialAmpere}
&\text{Amp\`{e}re-Maxwell:}&
-\nabla \cdot \bivec{b} &= \mu_0 \vec{J} + \mu_0\epsilon_0 \frac{\partial \vec{E}}{\partial t}
 \\
\label{eq:DifferentialFaraday}
&\text{Faraday's law:}&
\nabla \wedge \vec{E} &= -\frac{\partial \bivec{b}}{\partial t}
 \\
\label{eq:DifferentialMagGauss}
&\text{Magnetic Gauss:}&
\nabla \wedge \bivec{b} &= 0
\end{align}
(In two dimensions, $\rho \to \sigma$ and $\vec{J} \to \vec{K}$.)
As we'll see, in this form divergence always relates to source charges.

The derivatives of the electric field in Eqs.~\eqref{eq:DifferentialGauss} and~\eqref{eq:DifferentialFaraday} are mostly familiar.
Expressed in index notation, divergence and curl apply to a vector field $\vec{E}$ as follows:
\begin{align}
\label{eq:DivEIndex}
\nabla \cdot \vec{E} \quad &\to& \nabla \cdot \vec{E} &= \partial_i E_i
\quad,
\\
\label{eq:CurlEIndex}
\nabla \times \vec{E} \quad &\to& (\nabla \wedge \vec{E})_{ij}
 &= \partial_i E_j - \partial_j E_i
\quad.
\end{align}
The first of these is exactly the usual divergence: geometrically, ``how much is the vector field's magnitude changing along the vector's own direction?''
The second is the \defn{bivector curl} (or more formally, the \defn{exterior derivative}) of the vector field $\vec{E}$: geometrically, ``how is the vector field rotating?''%
\footnote{Another useful interpretation of the bivector curl is ``what sideways shift (perpendicular to the vector direction) increases the magnitude most?'' The result is a wedge product of the direction of fastest increase with the vector's rate of increase.}
It is well-defined in any number of dimensions, and in three dimensions where $\nabla \times \vec{E}$ is meaningful the two are interchangeable: just a bivector and its corresponding pseudovector.
(In the familiar fluid analogy, the $\nabla \wedge \vec{E}$ bivector field's orientation matches the fluid's actual circulating motion.)

\begin{figure}
\centering
\includegraphics{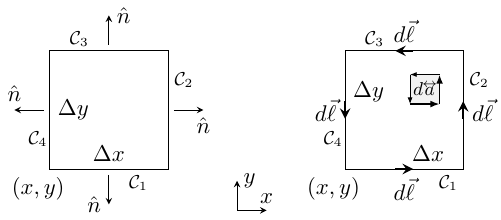} 
\caption{\label{fig:VecDerivatives}%
\textit{Left:} A small 2D Gaussian ``surface'' aligned with the coordinate axes, whose sides are curves $\mathcal{C}_1$--$\mathcal{C}_4$, each with its own normal vector. (In 3D this would be a cube with sides $\mathcal{A}_1$--$\mathcal{A}_6$.)
\textit{Right:} A small loop aligned with the $x$ and $y$ coordinate axes: its sides are four curves $\mathcal{C}_1$--$\mathcal{C}_4$ (this is the same in 2D or 3D). 
An (arbitrary) choice of direction for the boundary length elements $d\vec{\ell}$ is shown; the bivector area element $d\bivec{a}$ on the interior has matching orientation $\hat{x}\wedge\hat{y}$.} 
\end{figure}

For a more formal understanding, we can begin in two dimensions and apply the integral form of Gauss's law in Eq.~\eqref{eq:Gauss2D} to a very small area aligned with the coordinate axes as in Fig.~\ref{fig:VecDerivatives} (left).
If the area is small enough, the charge density will be approximately constant and can factor out of the integral: $\int_\mathcal{A} \sigma \,da \approx \sigma \, \Delta x \, \Delta y$.
The same applies to the integrals of $\vec{E}$ through each surface curve:
\begin{align}
\label{eq:Divergence2D}
\oint_{\mathcal{C}} \vec{E} \cdot \dvSurf 
 &= \int_{\mathcal{C}_1} \vec{E} \cdot (-\hat{y} \, dx)
   + \int_{\mathcal{C}_2} \vec{E} \cdot (\hat{x} \, dy)
\\
\nonumber
 &\phantom{{}={}}
   + \int_{\mathcal{C}_3} \vec{E} \cdot (\hat{y} \, dx)
   + \int_{\mathcal{C}_4} \vec{E} \cdot (-\hat{x} \, dy)
\\
\nonumber
 &\approx -E_y(\mathcal{C}_1) \,\Delta x
   + E_x(\mathcal{C}_2) \,\Delta y
\\
\nonumber
 &\phantom{{}={}}
  + E_y(\mathcal{C}_3) \,\Delta x
   - E_x(\mathcal{C}_4) \,\Delta y
\\
\nonumber
 &= \left( \frac{E_x(x+\Delta x, y) - E_x(x,y)}{\Delta x} \right.
\\
\nonumber
 &\qquad + \left. \frac{E_y(x, y+\Delta y) - E_y(x,y)}{\Delta y} \right) \, \Delta x \Delta y
\\
\nonumber
 &\approx \left( \partial_x E_x + \partial_y E_y \right) \, \Delta x \Delta y
 \quad.
\end{align}
The final term in parentheses matches Eq.~\eqref{eq:DivEIndex}.
In the second step, the notation ``$E_x(\mathcal{C}_2)$'' means ``the value of $E_x$ along the curve $\mathcal{C}_2$'' (which is approximately constant). 
The third step collects parallel terms and writes out the differences in the curve locations: $\mathcal{C}_1$ and $\mathcal{C}_3$ are the same except that $\mathcal{C}_3$ is shifted by $\Delta y$.
The final step recognizes the definition of the partial derivative (in the limit of small $\Delta x$ and $\Delta y$): the divergence is just the infinitesimal limit of ``flux per unit area.''
Altogether, this leads to the differential form of Gauss's law in Eq.~\eqref{eq:DifferentialGauss}.
The generalization to three dimensions is straightforward, integrating over surface planes instead of surface lines and factoring out $\Delta x \Delta y \Delta z$ to get ``flux per unit volume.''

\medskip

Meanwhile, the bivector curl of the vector field $\vec{E}$ emerges from applying the left side of Faraday's law, Eq.~\eqref{eq:Faraday}, to a tiny loop as in Fig.~\ref{fig:VecDerivatives} (right).
As before, for a small enough area the magnetic field $\bivec{b}$ is approximately constant: 
$\int_\mathcal{A} \tfrac{1}{2} \bivec{b} \dblcdot d\bivec{a}
 \approx \tfrac{1}{2} \bivec{b} \dblcdot (\Delta x \Delta y \, \hat{x}\wedge\hat{y})
 = b_{xy} \, \Delta x \Delta y$.
 As for the integrals along the edges,
 \begin{align}
\label{eq:BivecCurl2D}
\oint_\mathcal{C} \vec{E} \cdot d\vec{\ell}
 &= \int_{\mathcal{C}_1} \vec{E} \cdot (\hat{x} \, dx)
   + \int_{\mathcal{C}_2} \vec{E} \cdot (\hat{y} \, dy)
\\
\nonumber
 &\phantom{{}={}}
   + \int_{\mathcal{C}_3} \vec{E} \cdot (-\hat{x} \, dx)
   + \int_{\mathcal{C}_4} \vec{E} \cdot (-\hat{y} \, dy)
\\
\nonumber
 &\approx E_x(\mathcal{C}_1) \,\Delta x
   + E_y(\mathcal{C}_2) \,\Delta y
\\
\nonumber
 &\phantom{{}={}}
  - E_x(\mathcal{C}_3) \,\Delta x
  - E_y(\mathcal{C}_4) \,\Delta y
\\
\nonumber
 &= \left( \frac{E_y(x+\Delta x, y) - E_y(x,y)}{\Delta x} \right.
\\
\nonumber
 &\qquad - \left. \frac{E_x(x, y+\Delta y) - E_x(x,y)}{\Delta y} \right) \, \Delta x \Delta y
\\
\nonumber
 &\approx \left( \partial_x E_y - \partial_y E_x \right) \, \Delta x \Delta y
 \quad.
\end{align}
(In three dimensions, this is familiar as the $z$-component of $\nabla \times \vec{E}$, but it is very natural to just see it directly as encoding curl in the $xy$-plane.)
This exactly matches Eq.~\eqref{eq:CurlEIndex}, and it leads directly to Eq.~\eqref{eq:DifferentialFaraday}.
(Flux $\Phi \approx b_{xy} \,\Delta x\Delta y$.)
This will always be a two-dimensional calculation, but in three dimensions it can be repeated in any coordinate plane.
The bivector curl $\nabla \wedge \vec{E}$ can be understood as the ``infinitesimal circulation of $\vec{E}$ per unit area, oriented in the plane of maximum circulation.''

\bigskip

Derivatives for bivector magnetic fields are less familiar, but as with the vector derivatives above they follow naturally from the integral forms that we've seen. The curl of a pseudovector $\vec{B}$ is equivalent to the (negative) divergence of the corresponding bivector $\bivec{b}$:%
\footnote{For clarity, we always take the (single) dot product to mean ``contraction of adjacent tensor indices,'' even though $\partial_j b_{ij}$ may be more geometrically natural here than $\nabla \cdot \bivec{b} = \partial_j b_{ji}$.}
\begin{align}
\label{eq:DivBIndex}
\nabla \times \vec{B} \quad &\to& -(\nabla \cdot \bivec{b})_i &= -\partial_j b_{ji}
 = \partial_j b_{ij}
\quad.
\end{align}
The result has one tensor index: it is a vector quantity.

There are several helpful geometric interpretations of the divergence of a bivector.
One is to treat each component as an ordinary vector divergence.
For example, if you look at the bivector tiles in Fig.~\ref{fig:AmpereLine3D} from a point along the $z$-axis (projecting them edge-on onto the $xy$-plane), their ($+z$)-edge arrows spread out from the wire.
This is the $z$-component of the divergence: $\partial_j b_{zj}$.%
\footnote{Formally, we factor the bivector field into a part orthogonal to $z$ and a part that combines $\hat{z}$ with some vector in the $xy$-plane:
\begin{align}
\nonumber
\bivec{b} &= b_{xy} \, \hat{x}\wedge\hat{y} 
  + \hat{z} \wedge ( b_{zx} \hat{x} + b_{zy} \hat{y} )
\\
 &\equiv b_{xy} \, \hat{x}\wedge\hat{y} + \hat{z} \wedge \vec{v}^{\,(xy)}
 \quad.
\end{align}
Then the negative $z$-divergence of $\bivec{b}$ is just the ordinary divergence of $\vec{v}^{\,(xy)}$ (in the $xy$-plane): 
$-(\nabla\cdot\bivec{b})_z = -\partial_j b_{jz} = \partial_j v^{\,(xy)}_j$.}

\medskip

\begin{figure}
\centering
\includegraphics{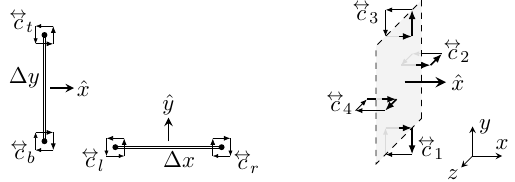} 
\caption{\label{fig:BivecDiv}%
\textit{Left:} Two Amp\`{e}rian ribbons aligned with the coordinate axes in 2D. Each endpoint's encircling unit bivector has orientation aligned to pass through the ribbon in the direction of the normal vector.
\textit{Right:} An Amp\`{e}rian loop in 3D, aligned parallel to the $yz$-plane. Each edge's encircling bivector orientation is aligned to agree with the normal vector $\hat{x}$.} 
\end{figure}

We can also build geometric intuition from the derivation of this differential formulation.
We'll start with Amp\`{e}re's law in two dimensions, Eq.~\eqref{eq:Ampere2D}, applied to ``Amp\`{e}rian ribbons'' extended in each coordinate direction as in Fig.~\ref{fig:BivecDiv}~(left).
Similar to the vector equations above, the current density is approximately constant:
$\int_\mathcal{C} \vec{K} \cdot \dvSurf \approx \vec{K} \cdot \hat{n} \,\Delta\ell$.
For the ribbon whose normal vector is $\hat{x}$ with endpoints $p_b$ and $p_t$, this becomes $K_x \Delta y$.
The left side for this ribbon reads
\begin{align}
\sum
\tfrac{1}{2} \bivec{b} \dblcdot \bivec{c}_p
 &= \tfrac{1}{2} \bivec{b}(p_b) \dblcdot \bivec{c}_b
   + \tfrac{1}{2} \bivec{b}(p_t) \dblcdot \bivec{c}_t
\\
\nonumber
 &= \tfrac{1}{2} \bivec{b}(y) \dblcdot (-\hat{x}\wedge\hat{y})
   + \tfrac{1}{2} \bivec{b}(y+\Delta y) \dblcdot (\hat{x}\wedge\hat{y})
\\
\nonumber
 &= -b_{xy}(y) + b_{xy}(y+\Delta y)
 \approx \partial_y b_{xy} \, \Delta y
\quad.
\end{align}
The second step used the encirclement bivectors $\bivec{c} = \hat{n} \wedge \hat{o}$ for each endpoint, and the final step used the definition of the partial derivative in the $\Delta y \to 0$ limit.
This last expression has only one term, but it matches Eq.~\eqref{eq:DivBIndex} because $b_{ij}$ is antisymmetric:
\begin{equation}
-\partial_j b_{jx} =
\partial_j b_{xj} = \partial_x b_{xx} + \partial_y b_{xy} = \partial_x (0) + \partial_y b_{xy}
\;.
\end{equation}

Similarly, for the ribbon whose normal vector is $\hat{y}$ with endpoints $p_l$ and $p_r$, 
the current density term in Amp\`{e}re's law becomes $K_y \Delta x$ while the sum on the left is
\begin{align}
\sum
\tfrac{1}{2} \bivec{b} \dblcdot \bivec{c}_p
\nonumber
 &= \tfrac{1}{2} \bivec{b}(x) \dblcdot (\hat{x}\wedge\hat{y})
   + \tfrac{1}{2} \bivec{b}(x+\Delta x) \dblcdot (-\hat{x}\wedge\hat{y})
\\
\nonumber
 &= b_{xy}(x) - b_{xy}(x+\Delta x)
 \approx -\partial_x b_{xy} \, \Delta x
\\
 &= -(\partial_x b_{xy} + \partial_y b_{yy}) \Delta x = -\partial_i b_{iy} \, \Delta x
\quad,
\end{align}
again matching the index notation.
Formally, each component is an ``encirclement of the boundary by $\bivec{b}$ per unit length (of the ribbon),'' so as a vector quantity, $-\nabla \cdot \bivec{b}$ is ``infinitesimal encirclement per unit length, oriented in the direction of maximum encirclement.''

This derivation leads to a second geometric interpretation of the divergence of a bivector. Similar to divergence for vector fields, we can phrase it as ``how much is the bivector field's magnitude increasing along the bivector's own plane?''
The vector direction points ``opposite the combined flow of the adjacent bivector tiles.'' 
For example, consider the magnetic field bivectors $\bivec{b}$ shown for two-dimensional current flow in Fig.~\ref{fig:AmpereLine2D}. Pick a point between two adjacent bivector tiles and add up the orientation arrows immediately above and below. If it is fully outside the flowing current, the two arrows have equal size and cancel out: $\nabla \cdot \bivec{b}=0$. But for a point inside the flowing current, the sum always points to the right, so $\nabla \cdot \bivec{b}$ points opposite that: ``upstream'' to the left.

\medskip

The same basic process works for an Amp\`{e}rian loop in three dimensions, as shown in Fig.~\ref{fig:BivecDiv}~(right) for a loop of side lengths $\Delta y$ and $\Delta z$ whose normal vector is $\hat{x}$, with edges $\mathcal{C}_1$--$\mathcal{C}_4$ have encircling bivectors $\bivec{c}_1$--$\bivec{c}_4$ as shown. The current density term is 
$\int_\mathcal{C} \vec{J} \cdot \dvSurf \approx \vec{J} \cdot \hat{n} \, da
= J_x \,\Delta y \, \Delta z$, and the encirclement integral along the boundary path is
\begin{align}
\nonumber
\oint_\mathcal{C} \tfrac{1}{2} \bivec{b} \dblcdot d\bivec{P}
 &= \int_{\mathcal{C}_1} \tfrac{1}{2} \bivec{b} \dblcdot (-\hat{x}\wedge\hat{y}) dz
  + \int_{\mathcal{C}_2} \tfrac{1}{2} \bivec{b} \dblcdot (-\hat{x}\wedge\hat{z}) dy
\\
\nonumber
  &\phantom{=} + \int_{\mathcal{C}_3} \tfrac{1}{2} \bivec{b} \dblcdot (\hat{x}\wedge\hat{y}) dz
  + \int_{\mathcal{C}_4} \tfrac{1}{2} \bivec{b} \dblcdot (\hat{x}\wedge\hat{z}) dy
\\
\nonumber
 &= -b_{xy}(\mathcal{C}_1) \Delta z - b_{xz}(\mathcal{C}_2) \Delta y
\\
\nonumber
 &\phantom{=} + b_{xy}(\mathcal{C}_3) \Delta z + b_{xz}(\mathcal{C}_4) \Delta y
\\
 &\approx (\partial_y b_{xy} + \partial_z b_{xz}) \, \Delta y \, \Delta z
  = \partial_j b_{xj} \, \Delta y \, \Delta z
\end{align}
This process is similar to the derivation of the ordinary divergence in Eq.~\eqref{eq:Divergence2D}. (This time we have not written out the definition of the derivatives leading to the final line.) There are corresponding calculations for $y$ and $z$.

Thus, the divergence of a bivector in three dimensions can be interpreted as ``encirclement of the boundary by $\bivec{b}$ per unit area (of the loop),'' and the vector quantity $-\nabla \cdot \bivec{b}$ is ``infinitesimal encirclement per unit area, oriented in the direction of maximum encirclement.''
The second geometric interpretation in terms of adjacent bivector tiles applies in three dimensions as well.
Using the magnetic field bivectors that circle around the axis of the three-dimensional current flow in Fig.~\ref{fig:AmpereLine3D} as an example, all of the inner edge orientations combine to point to the right, so as before $\nabla \cdot \bivec{b}$ points opposite that to the left.

\bigskip

Last but not least is the differential version of Gauss's law for magnetism. The divergence of a pseudovector $\vec{B}$ results in a ``pseudoscalar:'' a zero-dimensional quantity whose sign reverses when the coordinates are reflected. The corresponding operation for a bivector $\bivec{b}$ might be called the ``\defn{trivector curl}'' (another \defn{exterior derivative}):
\begin{align}
\label{eq:TrivecCurl}
\nabla \cdot \vec{B} \quad &\to& (\nabla \wedge \bivec{b})_{ijk} 
&= \partial_i b_{jk} + \partial_j b_{ki} + \partial_k b_{ij}
\;.
\end{align}
Because $b_{ij}$ is antisymmetric, you can check that this quantity is also antisymmetric. It is a \defn{trivector}: a quantity whose value has a right- or left-``handedness'' (rather than a simple positive or negative sign). The good news is that because magnetic monopoles are not known to exist, we do not need to concern ourselves much with how to interpret such things: it is simply zero.

In two-dimensional space that zero follows automatically from antisymmetry: at least two of the $ijk$ coordinate indices must match. 
In three-dimensional space there is only one independent component, $(\nabla \wedge \bivec{b})_{xyz}$: this equals the traditional pseudoscalar value.
Geometrically, $\nabla \wedge \bivec{b}$ can be interpreted as the rate at which the bivector field $\bivec{b}$ increases in a direction orthogonal to its own orientation plane: the absence of monopoles means it just \emph{doesn't}.

As with previous expressions, it is straightforward to derive Eq.~\eqref{eq:TrivecCurl} from the closed surface bivector flux of Eq.~\eqref{eq:GaussMag}. 
We will just sketch this, in parallel with previous derivations.
Begin from a tiny cubic region as shown in Fig.~\ref{fig:MagneticGauss} and integrate over each  of the six faces:
\begin{align}
\nonumber
\oint_\mathcal{A} \tfrac{1}{2} \bivec{b} \dblcdot d\bivec{a}
 &=  b_{yz} (\mathcal{A}_\text{right}) \Delta y \, \Delta z
       - b_{yz} (\mathcal{A}_\text{left}) \Delta y \, \Delta z
\\
\nonumber
 &\phantom{=} + b_{zx} (\mathcal{A}_\text{top}) \Delta z \, \Delta x
       - b_{zx} (\mathcal{A}_\text{bottom}) \Delta z \, \Delta x
\\
\nonumber
 &\phantom{=} + b_{xy} (\mathcal{A}_\text{front}) \Delta x \, \Delta y
       - b_{xy} (\mathcal{A}_\text{back}) \Delta x \, \Delta y
\\
 &\approx (\partial_x b_{yz} + \partial_y b_{zx} + \partial_z b_{xy}) \, \Delta x \, \Delta y \, \Delta z
\end{align}
The final result for total magnetic flux is the component $(\nabla \wedge \bivec{b})_{xyz}$ times the volume of the cube.


\section{Pedagogy and Conclusions}
\label{sec:conclusions}

The key takeaway messages of all this depend on what approach a given instructor adopts in the classroom.
For those who choose to fully embrace the bivector formalism, the main lesson is that the formalism extends to all of electrodynamics and that two-dimensional examples can be a good way to build understanding before extending to three dimensions.
It also keeps symmetry arguments front and center in all of Maxwell's equations.

Both forms of geometric intuition for bivector divergence in Amp\`{e}re's law can be seen as ``natural.'' 
The analogy of the boundary's encircling bivectors as ``water wheels'' is highly visual: an instructor could show iron filings following circular field lines around current in a wire, and then invite the class to imagine tiny water wheel ``beads'' strung around the field line spinning as if pushed by the current inside.
There is also a direct parallel between ``electric field vectors spread away from source charges'' (vector divergence) and ``the edges of magnetic bivector tiles aligned along a source current spread out from the current'' (bivector divergence).
The traditional ``magnetic field vectors point in circles around currents'' does not have a comparably natural justification.


But as discussed in previous work,\cite{Jensen:2023mag} adopting the full bivector formalism is a big leap: many instructors will prefer the smaller step of representing the magnetic field with ``decorated vectors:'' a vector arrow with a small curved arrow around its middle that shows the corresponding right-hand rule orientation.
This means missing out on the opportunity for simple two-dimensional examples, but it is compatible with standard textbook content and many of the other benefits of the bivector description are still partially available.

Students can still use symmetry arguments to reason about magnetic field directions, as long as the instructor makes it clear that only the curved ``decoration'' arrow's reflection can be trusted. The visual metaphor of ``water wheels'' can motivate the right-hand rule direction of the pseudovector field around a field line.
And this approach has the advantage that students could get by learning just the traditional divergence and curl without also having to learn their bivector equivalents.

One cost of the ``decorated vector'' middle ground is that only the pure bivector approach can generalize to \emph{higher} dimensions, such as relativistic spacetime. 
The electromagnetic field tensor $\fourvec{F}^{\mu \nu}$ is a bivector with a natural interpretation in terms of ``tiles'' in four-dimensional spacetime,\cite{Jensen:2023mag} and (as will be discussed in future work) the geometric concepts introduced here apply just as well in that context. Bivector encirclement of a boundary and bivector flux along a two-dimensional surface will allow us to illustrate Maxwell's equations in relativistic form. 
But introducing the concept of decorated vectors can at least plant the seed of the bivector description: these more advanced ideas won't come out of nowhere.

\section*{Acknowledgments}

Thanks to Alma College for a sabbatical supporting this work. Figures~\ref{fig:LoopFieldReflection}--\ref{fig:BivecDotProds} previously appeared in Ref.~\cite{Jensen:2023mag}.

\bibliography{bivectors}

\end{document}